\documentclass[conference]{IEEEtran}
\IEEEoverridecommandlockouts
\usepackage{cite}
\usepackage{amsmath,amssymb,amsfonts}
\usepackage{algorithmic}
\usepackage{graphicx}
\usepackage{textcomp}
\usepackage{xcolor}
\usepackage{amsmath}
\usepackage{amssymb}
\usepackage{booktabs}
\usepackage{siunitx}
\usepackage{tabularx}
\usepackage{array}
\def\BibTeX{{\rm B\kern-.05em{\sc i\kern-.025em b}\kern-.08em
    T\kern-.1667em\lower.7ex\hbox{E}\kern-.125emX}}
\begin{document}

\title{How Much Precision Does FHRR Need? Hardware-Efficient Vector Symbolic Computing with Quantized Phases}

\author{
\IEEEauthorblockN{
    Shay Snyder$^{1,\dagger}$, Hamed Poursiami$^1$, Maryam Parsa$^{1,\dagger}$}
\IEEEauthorblockA{
    $^1$George Mason University, Fairfax, United States\\
    $^\dagger$Corresponding Authors: \{ssnyde9,mparsa\}@gmu.edu}
}

\maketitle

\begin{abstract}
Fourier Holographic Reduced Representations (FHRR) encode symbols of unit phasors, however conventional implementations store and manipulate them with complex floating-point arithmetic.
We investigate how much precision FHRR requires through quantized-phase FHRR (qFHRR) that represents each component as an $N$-bit integer phase.
We compare qFHRR against complex, Cartesian, and modular composite representations across associative retrieval, factorization, and path integration tasks.
At matched dimensionality, quantization introduces a predictable capacity penalty proportional to $\operatorname{sinc}^{2}(\pi/2^{N})$.
However, under equal storage budgets, 2-bit qFHRR stores $\approx26\times$ more key--value associations than complex FP32.
Resonator factorization is more sensitive to precision, and path integration loses displacements smaller than one phase bin regardless of dimensionality.
Because these storage advantages do not translate into faster execution when phase indices are decoded into floating point for standard operations, we introduce HyPack, a hybrid-precision packed qFHRR implementation that combines 2-bit Gray-coded phase storage and native bitwise operations with wider integer accumulation.
Compared with complex FP32 and Cartesian FP16 fused kernels, HyPack retrieves $3.3\times$ faster per query with $4.3\times$ less energy on a CPU and $1.4\times$ faster with $2.6\times$ less energy on a GPU.
In path integration, its integer kernel matches fused floating-point integration on the CPU, while its $32\times$ smaller state and codebook cut the cost of decoding. Therefore, integration with decoding is $4.4\times$ faster with $5.0\times$ less energy on the CPU and $1.3\times$ faster on the GPU.
Separating representation precision from intermediate precision therefore enables more efficient phase-based vector symbolic computation on conventional processors.
Code will be released upon publication.
\end{abstract}


\section{Introduction}
\label{sec:intro}

\noindent
Vector Symbolic Architectures (VSAs), also known as hyperdimensional computing, represent information with high-dimensional vectors and a set of algebraic operations for binding, bundling, and similarity~\cite{gayler2003vsa,kanerva2009hdc,kleyko2022survey1,kleyko2023survey2}.
Their distributed representations are robust to noise, and their algebra supports compositional and interpretable computation\cite{rahimi2016hdc,karunaratne2020inmemory}.
Among the various VSA families~\cite{schlegel2022comparison}, Fourier Holographic Reduced Representations (FHRR)~\cite{plate1995hrr,plate2003book} represent symbols as vectors of $D$ unit phasors.
VSA algebra supports associative memory, resonator networks for factorizing composite representations~\cite{frady2020resonator1,kent2020resonator2}, and fractional power encoding of continuous values~\cite{komer2019fractional,frady2021vfa}.

Despite leveraging phase-based algebra, conventional FHRR implementations rely on complex floating-point arithmetic~\cite{heddes2023torchhd} to represent a phasor with a single degree of freedom.
Spiking phasor networks offer a different perspective by encoding phase through spike timing relative to a reference oscillation~\cite{frady2019rhythmic,orchard2023spikingphasor}, and have demonstrated FHRR resonator networks on neuromorphic hardware~\cite{renner2024neuromorphic} such as Loihi~\cite{davies2018loihi}.
These results separate the information FHRR represents from numerical implementation, but they leave open how much phase resolution is necessary to preserve FHRR's computational properties without complex arithmetic.

Reducing phase resolution trades representation fidelity for storage efficiency.
At fixed dimensionality, quantization introduces phase error that degrades similarity measurements, whereas fixed memory budgets and fewer bits per component can permit higher-dimensional representations that recover or exceed the lost capacity~\cite{frady2018sequence,clarkson2023capacity}.
How favorable this tradeoff is depends on the task.
Associative retrieval must tolerate interference between stored items~\cite{kleyko2022survey1}, resonator networks rely on iterative cleanup to recover composite factors~\cite{frady2020resonator1}, and continuous encodings must preserve small phase increments that quantization may discard~\cite{frady2021vfa}.
These considerations raise the central question of this work: \emph{how should representation precision, intermediate arithmetic, and dimensionality be balanced to maximize capacity and computational efficiency?}

We address this question through quantized-phase FHRR (qFHRR), which represents each phasor as an $N$-bit integer phase index, where binding becomes modular integer addition and similarity and bundling operate on quantized values.
We compare qFHRR against complex floating-point, Cartesian fixed-point, minifloat, and modular composite representations~\cite{snaider2014mcr} on associative key--value retrieval, resonator-network factorization, and spatial path integration.
Accuracy-matched benchmarks on CPU, GPU, and embedded platforms then show that reduced storage rarely lowers latency or energy, because standard quantized implementations decode phase indices into floating-point tensors before computing similarity or leverage inefficient LUT alternatives.
We therefore introduce \emph{HyPack}, a hybrid-precision packed qFHRR that stores two-bit phases as Gray-coded bit planes where binding, inversion, and similarity reduce to native bitwise operations and population counts.
HyPack retains wider integer accumulators for operations such as path integration before quantizing to a two-bit readout.
Because fusing framework operations into one kernel accelerates every format, we compare HyPack against complex fused FP32 and Cartesian FP16 kernels.

In summary, the major contributions of this work are:
\begin{itemize}
    \item We systematically compare phase precisions and show empirically that the matched-$D$ capacity penalty follows $\operatorname{sinc}^{2}(\pi/2^{N})$.
    \item We show that trading precision for dimensionality under a fixed storage budget increases associative capacity. For example, 2-bit qFHRR stores about $26\times$ more key--value associations than complex FP32 at 16\,Kibit.
    \item We identify task-dependent precision requirements where low precision inhibits resonator convergence, and path integration loses sub-bin phase increments that added dimensions cannot recover.
    \item We show that reduced storage alone does not improve execution efficiency, since decoding phase indices to floating point or leveraging LUTs can make qFHRR up to $3.8\times$ slower than complex FP32.
    \item We introduce HyPack, a hybrid-precision qFHRR, and compare it against equally fused complex FP32 and Cartesian FP16 kernels. On a CPU, HyPack retrieves $3.3\times$ faster per query with $4.3\times$ less energy and completes path integration $4.4\times$ faster with $5.0\times$ less energy. On a GPU, HyPack retrieves $1.4\times$ faster per query with $2.6\times$ less energy and completes path integration $1.3\times$ faster with $1.3\times$ less energy.
\end{itemize}

% NICE 2027 background revision.
% One complete sentence per LaTeX source line.

\section{Background and Related Work}
\label{sec:background}

\subsection{Phase-Based Vector Symbolic Computing}

\noindent
FHRR encode symbols as $D$-dimensional hypervectors of unit phasors $z_d=e^{\mathrm{i}\theta_d}$~\cite{plate1995hrr,plate2003book}.
FHRR and broader VSA algebra underlies resonator networks, which recover bound factors by iterative search~\cite{frady2020resonator1,kent2020resonator2}, and spatial semantic pointers, which encode continuous locations through fractional powers~\cite{komer2019fractional,frady2021vfa}.
Related encodings have been applied to probabilistic mapping~\cite{snyder2026vsaogm}, graph learning~\cite{poursiami2026vs}, path integration~\cite{krausse2025grid}, and simultaneous localization and mapping~\cite{dumont2023exploiting}.
These applications stress different aspects of the algebra, since associative retrieval depends on similarity measurements that survive interference from stored items, resonator networks extend this dependence to iterative factorization, and path integration requires small changes in position to remain distinguishable after quantization.

\subsection{Numerical and Neuromorphic Implementations of FHRR}

\noindent
Conventional implementations such as Torchhd store each phasor as a complex floating-point value~\cite{heddes2023torchhd}.
An equivalent Cartesian representation separates the real and imaginary parts into a real-valued vector $\tilde{\mathbf{z}}=[\operatorname{Re}(\mathbf{z});\operatorname{Im}(\mathbf{z})]\in\mathbb{R}^{2D}$ that preserves the FHRR algebra while allowing fixed-point or reduced-precision floating-point arithmetic.
Related real-valued sinusoidal encodings bind by Hadamard product~\cite{schlegel2026realvalued}.

Discrete-phase representations replace complex storage with an integer phase index.
Modular composite representations (MCR) encode each component in $\mathbb{Z}_r$ with binding $(\mathbf{a}\odot\mathbf{b})_d=(a_d+b_d)\bmod r$, but measure similarity with a circular Manhattan distance rather than cosine similarity~\cite{snaider2014mcr}.
MCR approaches complex-valued capacity with reduced storage, has been realized on a dedicated FPGA accelerator~\cite{angioli2025mcr}, and related cyclic-group VSAs combine modular binding with cosine similarity~\cite{yu2022cgr}.
Our qFHRR formulation adopts this discrete phase algebra with $N$-bit indices in $\mathbb{Z}_{2^N}$ while preserving FHRR's cosine similarity through lookup tables indexed by phase differences.
Rather than introducing a new binding algebra, we ask how phase resolution affects FHRR operations, capacity, and execution on general-purpose hardware.
Spike-timing representations go a step further by encoding phase relative to a reference oscillation~\cite{frady2019rhythmic,orchard2023spikingphasor,orchard2024spikingphasors}, and spiking resonator networks have demonstrated visual scene understanding on neuromorphic hardware~\cite{renner2024neuromorphic}.

\subsection{Hardware-Aware Precision Tradeoffs}

\noindent
Previous studies characterized the representational properties and theoretical capacity of VSA models~\cite{schlegel2022comparison,clarkson2023capacity}, dedicated MCR and FHRR accelerators have demonstrated the potential of phase arithmetic in hardware~\cite{angioli2025mcr,martino2026fhrr}, and GPU frameworks such as OpenHD and HDTorch have accelerated VSA workloads on general-purpose processors~\cite{kang2022openhd,simon2022hdtorch}.
None of these studies address how phase precision jointly influences task accuracy and computational efficiency across conventional processor architectures, where the outcome depends on memory layout, packing overhead, vectorization, and native support for the required operations.
These tradeoffs are the subject of the comparison that follows.

\begin{table*}[t]
  \centering
  \caption{Numerical representations evaluated in this work.
  Bits per dimension count stored atom values only. For the index
  formats, $K=2^N$ and $\tilde q\in[-K/2,K/2)$ is the centered
  representative of the stored index $q$.}
  \label{tab:formats}
  \footnotesize
  \setlength{\tabcolsep}{3.5pt}
  \renewcommand{\arraystretch}{1.05}
  \begin{tabularx}{\textwidth}{@{}l >{\raggedright\arraybackslash}p{2.35cm} >{\raggedright\arraybackslash}p{2.2cm} *{4}{>{\raggedright\arraybackslash}X}@{}}
    \toprule
    \textbf{Format} & \textbf{Variants (bits/dim)} & \textbf{Stored per dim}
      & \textbf{Bind} & \textbf{Similarity} & \textbf{Bundle projection}
      & \textbf{Fractional power} \\
    \midrule
    Complex FP
      & FP32 (64)
      & $(\mathrm{Re},\mathrm{Im})$ interleaved
      & complex multiply
      & complex dot/GEMM, normalize
      & $z/|z|$
      & $\operatorname{atan2}$, scale, $\sin/\cos$ \\
    Cartesian float
      & FP32 (64), FP16 (32)
      & $(\mathrm{Re},\mathrm{Im})$ planes
      & 4 mul + 2 add
      & real GEMM over $2D$, normalize
      & $z/|z|$
      & $\operatorname{atan2}$, scale, $\sin/\cos$ \\
    Cartesian fixed
      & int16 (32), int8 (16), int4 (8)
      & $(\mathrm{Re},\mathrm{Im})\cdot s$, $s=2^{w-1}{-}1$
      & product, requantize, saturate
      & integer dot/GEMM, normalize
      & $\operatorname{atan2}$, $\sin/\cos$, requantize
      & $\operatorname{atan2}$, scale, $\sin/\cos$, requantize \\
    Cartesian minifloat
      & E4M3 (16), E5M2 (16), E2M1 (8)
      & $(\mathrm{Re},\mathrm{Im})$ codes
      & upcast, multiply, round
      & upcast, real dot/GEMM, normalize
      & upcast, project, round
      & upcast, $\operatorname{atan2}$, scale, $\sin/\cos$, round \\
    % Floating-point angle
    %   & FP32 (32), FP16 (16)
    %   & $\theta\in[-\pi,\pi)$
    %   & add, wrap
    %   & mean of $\cos(\theta_a{-}\theta_b)$
    %   & decode, add, $\operatorname{atan2}$
    %   & multiply, wrap \\
    MCR~\cite{snaider2014mcr,angioli2025mcr}
      & $N=2$--6, 8 ($N$)
      & index $q\in\mathbb{Z}_{2^N}$
      & $(a+b)\bmod K$
      & circular $L_1$
      & decode, add, round; zero sum: rounded mean index
      & $\operatorname{round}(p\tilde q)\bmod K$ \\
    \textbf{qFHRR (ours)}
      & $N=1$--6, 8, 16 ($N$)
      & index $q\in\mathbb{Z}_{2^N}$
      & $(a+b)\bmod K$
      & cosine LUT, reduce
      & decode, add, round to bin
      & $\operatorname{round}(p\tilde q)\bmod K$ \\
    \bottomrule
  \end{tabularx}
\end{table*}

\begin{figure*}[t]
  \centering
  \includegraphics[width=0.80\textwidth]{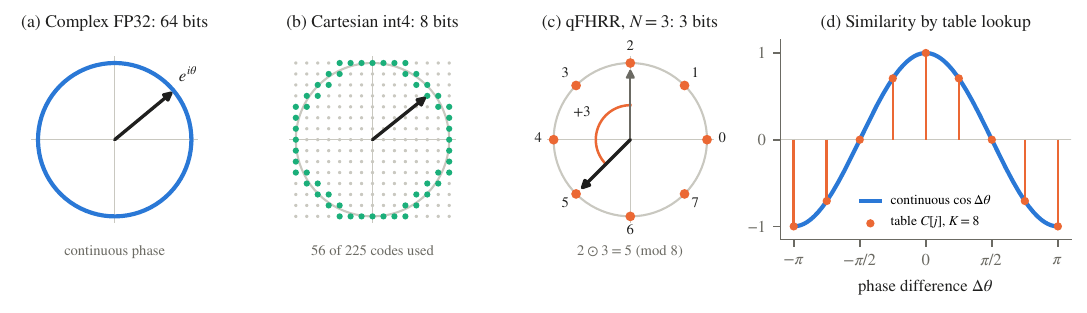}
  \caption{Numerical representations of a unit phasor.
  (a) Complex FP32: 64 bits for an unrestricted point on the unit circle.
  (b) Cartesian int4: 8 bits for two components on a $15\times15$ grid, only the points nearest the circle are reached by quantizing unit phasors.
  (c) Three-bit qFHRR: eight phase indices with modular binding $2\odot3=5\pmod{8}$.
  (d) Similarity samples the cosine from a $K=8$ lookup table.}
  \label{fig:formats}
\end{figure*}

\section{Research Methods}
\label{sec:methods}

\noindent
We compare numerical implementations of FHRR that differ in their encoding, intermediate arithmetic, and treatment of finite-precision results.
Table~\ref{tab:formats} summarizes the formats and Fig.~\ref{fig:formats} illustrates their constellations.
Task-specific procedures are presented with their results in Section~\ref{sec:results}.

\subsection{Shared FHRR Algebra}

\noindent
An FHRR atom is a $D$-dimensional vector of unit phasors, $\mathbf{z}=e^{i\boldsymbol{\theta}}$, with independently sampled phases $\boldsymbol{\theta}\sim\mathcal{U}[-\pi,\pi)^D$.
The FHRR algebra comprises five operations:
\begin{align}
  \text{Binding:}\quad
  &(\mathbf{a}\odot\mathbf{b})_d = a_d b_d,
  \label{eq:fhrr-bind}\\
  \text{Inversion:}\quad
  &(\mathbf{a}^{-1})_d = \overline{a_d},
  \label{eq:fhrr-inverse}\\
  \text{Similarity:}\quad
  &\operatorname{sim}(\mathbf{a},\mathbf{b})
  =\frac{\operatorname{Re}\sum_d\overline{a_d}b_d}
  {\lVert\mathbf{a}\rVert_2\lVert\mathbf{b}\rVert_2},
  \label{eq:fhrr-sim}\\
  \text{Bundling:}\quad
  &\bigoplus_{j=1}^{m}\mathbf{a}^{(j)}
  = \Pi\!\left(\sum_{j=1}^{m}\mathbf{a}^{(j)}\right),
  \label{eq:fhrr-bundle}\\
  \text{Fractional power:}\quad
  &(\mathbf{a}^{p})_d = e^{ip\theta_{a,d}}.
  \label{eq:fhrr-power}
\end{align}
Here $\Pi(z)=z/|z|$ projects each nonzero component of a sum onto the unit circle.
For unit phasors, similarity reduces to the mean cosine of the componentwise phase differences, and fractional powers use the phases $\theta_{a,d}\in[-\pi,\pi)$ to support continuous encodings~\cite{komer2019fractional,frady2021vfa}.
The comparison formats approximate these operations with different stored values and arithmetic, so format-specific similarity and bundling rules are identified below.
For paired comparisons, each seed fixes the random atoms, inputs, and reference trajectories.
% All accuracy experiments use a reference implementation that evaluates each operation in double precision and rounds intermediate results into the comparison format.
% The hardware benchmarks use a separate PyTorch implementation that accumulates bundles in FP32 and in int32 arithmetic for fixed-point formats.

\subsection{Quantized-Phase FHRR}
\label{sec:qfhrr}

\noindent
qFHRR stores each phasor as an $N$-bit index $q_d\in\mathbb{Z}_K$ with $K=2^N$, where index $q_d$ represents phase $2\pi q_d/K$ and a continuous phase is encoded as
\begin{equation}
  q_d=\operatorname{round}\!\left(\frac{K\theta_d}{2\pi}\right)
       \bmod K.
  \label{eq:qencode}
\end{equation}
Binding and inversion become modular integer operations, similarity is evaluated from the difference between phase indices, and fractional powers scale the centered representative $\tilde a_d\in[-K/2,K/2)$ of the index:
\begin{align}
  (\mathbf{a}\odot\mathbf{b})_d
    &= (a_d+b_d)\bmod K, \label{eq:qbind}\\
  (\mathbf{a}^{-1})_d
    &=(-a_d)\bmod K, \label{eq:qinv}\\
  \operatorname{sim}_{q}(\mathbf{a},\mathbf{b})
    &= \frac{1}{D}\sum_{d=1}^{D}
       C[(a_d-b_d)\bmod K], \label{eq:qsim}\\
  (\mathbf{a}^{p})_d
    &=\operatorname{round}(p\tilde a_d)\bmod K.
    \label{eq:qpow}
\end{align}
Here $C[j]=\cos(2\pi j/K)$ is a $K$-entry lookup table.
% Because $K$ is a power of two, modulo reduction is a bit mask, and for $N=8$ unsigned 8-bit overflow provides it automatically, so binding and inversion are exact within the discrete phase group and never decode phases into complex values.
Similarity replaces complex products with index differences, table accesses, and a reduction. 
% but conventional processors offer no native instruction for this gather-and-reduce, so the framework implementation in Section~\ref{sec:results} decodes both operands to Cartesian FP32 through $K$-entry cosine and sine tables and evaluates~\eqref{eq:qsim} as an FP32 matrix multiplication.

For $K=4$ the table is $C=(1,0,-1,0)$.
Mapping each index to a Gray-coded pair of signs $(s_1,s_2)\in\{\pm1\}^2$ gives $C[(a_d-b_d)\bmod 4]=\tfrac{1}{2}(s_{1,a}s_{1,b}+s_{2,a}s_{2,b})$, so~\eqref{eq:qsim} is evaluated exactly by storing a vector as two bit planes and computing $D$ minus the population count of their exclusive-or.
Under the same map, binding~\eqref{eq:qbind} is bitwise logic on the planes and inversion~\eqref{eq:qinv} swaps them.
Exclusive-or and population count are native instructions on conventional CPUs and GPUs, so this kernel, which we call HyPack, requires no dedicated accelerator.
% HyPack is therefore a kernel for one value of $K$ rather than a different representation, since for $K\geq 8$ the table contains entries other than $0$ and $\pm1$ and similarity requires a gather or a multiplication.
HyPack is therefore a kernel for one value of $K$ rather than a different representation.
Because two-bit qFHRR similarity is the mean of two bipolar similarities, the HyPack kernel is the Hamming-similarity kernel of binary VSAs such as binary spatter codes~\cite{kleyko2022survey1}.
What qFHRR adds is modular phase binding with fractional powers of a wide base, which HyPack uses for path integration before reading out to this binary-compatible format.

% Fractional powers and bundling introduce quantization error because a scaled phase may fall between discrete levels and bundling accumulates the sine and cosine values before quantizing to the nearest bin.
% Two cases require a tie rule: a resultant magnitude at most $10^{-9}$ of the total bundle weight is treated as zero, and a projected phase within $10^{-9}$ bins of a bin midpoint as an exact half-bin tie.
% Both are resolved with a tie vector, a random atom from a dedicated seed stream bound with the product of the bundle arguments and cyclically shifted by one dimension, so that tied components are deterministic but independent across bundles.
% A zero component takes the tie vector's phase index, and a half-bin tie takes whichever candidate bin is circularly nearer to it, avoiding the even-index bias of round-half-to-even.

For uniformly distributed phases, quantization introduces an error $\varepsilon\sim\mathcal{U}[-\pi/K,\pi/K)$, and the expected agreement between a continuous phasor and its quantized counterpart is
\begin{equation}
  \mathbb{E}[\cos\varepsilon]
  = \operatorname{sinc}\!\left(\frac{\pi}{K}\right)
  = \frac{\sin(\pi/K)}{\pi/K}.
  \label{eq:quant-attenuation}
\end{equation}
For two independently quantized phases, the expected agreement relative to their continuous phase difference decreases by $\operatorname{sinc}^{2}(\pi/K)$.

\subsection{Comparison Formats}

\noindent
\textit{Complex floating point.}
The reference stores each phasor as two FP32 values, 64 bits per component, consistent with TorchHD~\cite{heddes2023torchhd}, binds by complex multiplication, and serves as the accuracy baseline throughout.

\textit{Cartesian representations.}
These store the real and imaginary parts as a real-valued tensor $\tilde{\mathbf z}\in\mathbb R^{2D}$, compute similarity with real dot products or GEMM, and bind through complex multiplication in real arithmetic.
We evaluate FP32, FP16, signed fixed point with $w\in\{16,8,4\}$ bits, and the minifloats E4M3, E5M2, and E2M1.
ExMy denotes $x$ exponent and $y$ fraction bits plus a sign bit.
Fixed-point values use a scale $s=2^{w-1}-1$ and are rounded to integers in $[-s,s]$, with products rescaled, rounded, and saturated before storage. Minifloats are stored at their specified precision and upcast for computation when native arithmetic is unavailable.
% Similarity is normalized to account for quantization-induced changes in magnitude.

% \textit{Floating-point angle.}
% These store each phase $\theta_d$ directly in FP32 or FP16, bind by adding angles and wrapping to $[-\pi,\pi)$, scale angles for fractional powers, and compute similarity from the cosine of phase differences.

\textit{Modular composite representations.}
MCR uses the same $N$-bit indices and modular binding as qFHRR but measures similarity by the shortest circular distance between indices~\cite{snaider2014mcr,angioli2025mcr}:
\begin{align}
  \delta(\mathbf a,\mathbf b)
    &= \sum_{d=1}^{D}
    \min\bigl(|a_d-b_d|,K-|a_d-b_d|\bigr),
  \\
  \operatorname{sim}_{\mathrm{MCR}}(\mathbf a,\mathbf b)
    &= 1-\frac{4\delta(\mathbf a,\mathbf b)}{KD}.
  \label{eq:mcr-sim}
\end{align}
This similarity is one for identical vectors and has an expected value of zero for independent uniform indices.

\subsection{Evaluation Protocol}
\label{sec:protocol}

\noindent
We use two accuracy comparisons to separate the effects of numerical precision from the benefits of increased dimensionality.
At \emph{matched dimensionality}, all formats use $D\in\{256,512,1024,2048,4096,8192\}$.
At \emph{matched atom storage}, each format receives $B=16$\,Kibit and $D=\lfloor B/b\rfloor$, where $b$ is the stored bit width per dimension.
Here, complex FP32 supports $D=256$ and 1-bit qFHRR supports $D=16{,}384$.
% The budget counts only packed atom data, with lookup tables and temporary buffers accounted for separately in the hardware evaluation.
Accuracy is reported over 20 shared random seeds with 95\% percentile-bootstrap confidence intervals.

% Hardware benchmarks instead compare \emph{accuracy-matched} configurations.
% For each workload and format, we select the smallest dimensionality $D^\ast\in\{1024,4096\}$ that meets an accuracy threshold on the target device, since memory scales with $D$ and latency differences between eligible dimensionalities fall within run-to-run variation.
% Retrieval stores 128 key--value pairs, scores each query against 1024 candidate values, and must reach at least 99\% of the top-1 accuracy of the complex FP32 reference at the same load.
% Path integration applies 1000 displacements of 0.05 kernel lengths, and the similarity of the final state to the one-shot encoding of the endpoint must reach at least 99\% of the complex FP32 value.
% The reference is complex FP32 at $D=4096$ measured on the same device.
% Factorization is excluded because its runtime is dominated by the data-dependent number of resonator iterations rather than by the cost of individual operations.
% Each format is evaluated on 10 tasks, one per seed, after two untimed warmup tasks, with eligibility decided on the same seeds used to report latency and energy.
% Every eligible configuration clears its threshold by a wide margin except Cartesian FP16 path integration, whose final similarity of 0.9898 falls just below 0.99 and is therefore reported as infeasible.

We benchmark an AMD Ryzen 7 3700X CPU, an NVIDIA RTX 3090 GPU, and an NVIDIA Jetson Orin Nano Super.
% Phase-index formats are timed in two implementations.
% Decoded entries convert phases to Cartesian values and score them by matrix multiplication, with the candidate codebook decoded inside every call so that decoding cost is part of the reported latency.
Implemented in C++/OpenMP on the CPU and Triton on the GPU, the 2-bit HyPack entries use the bit-plane kernel of Section~\ref{sec:qfhrr}, and complex FP32 and Cartesian FP16 are additionally measured as fused kernels that apply the unbind inside the similarity kernel and all path steps in one call.
The C++ kernels are compiled with \texttt{-O3 -march=native}; the fused baselines use AVX2/FMA intrinsics on the 3700X but generic C++ loops on the Orin CPU, and on the GPUs the fused FP16 kernel uses tensor cores while the fused complex FP32 kernel uses IEEE FP32 arithmetic.
% On the GPUs each HyPack call is one Triton launch, whereas on the CPUs its unbind uses PyTorch bitwise operations before the popcount kernel.
Every reported measurement includes data conversion, unpacking, lookup tables, and reductions.
A retrieval call unbinds all 128 stored keys and scores each against the 1024 candidates. We report its time and energy divided by 128, i.e., the per-query cost within a 128-query batch rather than single-query latency.
A path call integrates one trajectory, covering all updates and the final readout.
Path integration is additionally timed with a decode stage that scores the final state against 701 candidate positions along the path axis, spaced 0.01 kernel lengths apart, encoded and compared in the format's own storage and similarity.

Time is the median over 10 tasks, one per seed, after two untimed warm-up tasks; each task is a single call with device synchronization.
CPU energy is measured with the RAPL package-domain counter from the Linux powercap interface, GPU energy on the RTX 3090 is measured with the NVML board energy counter, and the Jetson uses its INA3221 power monitor.
Energy is accumulated over at least 2~s of repeated calls to the timed region and reported per call.
Idle power is measured for 10~s after a 30~s cooldown at the start and end of each run.
Because the start-of-run CPU reading can include residual activity from process launch (up to 18.0~W, against 12.7--16.4~W at the end), CPU energies subtract the end-of-run idle power, whereas GPU energies subtract the start-of-run value, with start and end readings within 2.4~W.
The CPU runs with the performance governor and boost enabled on 8 threads pinned to logical CPUs 0--7, and the GPU runs in persistence mode with a 390~W board power limit.
The Jetson runs in its 25~W power mode with \texttt{jetson\_clocks} holding the CPU, GPU, and memory clocks at 1.73, 1.02, and 3.2~GHz on 6 threads pinned to CPUs 0--5.
Its energy is read from the INA3221 monitor on the module's VDD\_CPU\_GPU\_CV rail.

\begin{figure*}[t]
  \centering
  \includegraphics[width=0.95\textwidth]{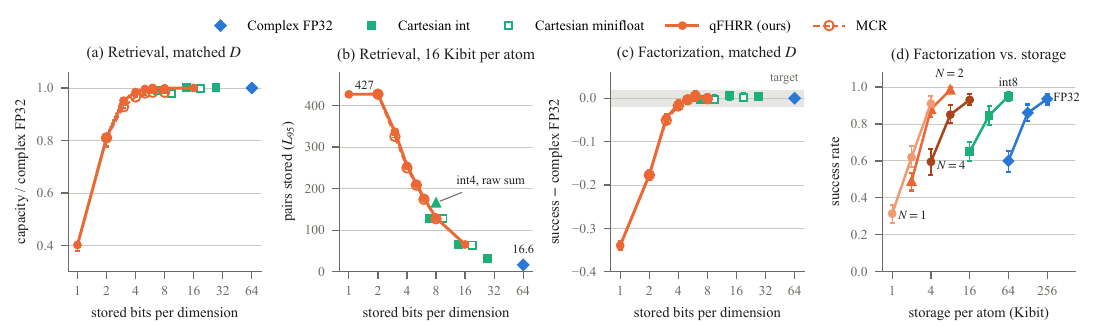}
  \caption{Precision--dimensionality tradeoffs in retrieval and factorization.
    (a)~Retrieval capacity relative to complex FP32 at matched $D$; markers are means over $D=256$--$8192$ and bars span the range over $D$, with the dotted curve showing the prediction $\operatorname{sinc}^{2}(\pi/K)$.
    (b)~Capacity $L_{95}$ with 16~Kibit atoms, $D=\lfloor 16384/b\rfloor$; the triangle is an unnormalized Cartesian int4 memory.
    (c)~Seed-paired factorization success relative to FP32 over the 26 configurations where FP32 success lies between 5\% and 95\%, with the shaded band marking the two-percentage-point criterion.
    (d)~Factorization success versus packed atom storage for clean composites with $F=3$, $M=128$, and $D=1024$--$4096$.
    In (b)--(d), error bars are 95\% bootstrap confidence intervals over 20 seeds. Cartesian FP32 is omitted because it reproduced complex FP32. Cartesian FP16 is omitted because it was within 0.1\% of complex FP32 at matched $D$.}
  \label{fig:res-tradeoff}
\end{figure*}

\section{Results \& Discussion}
\label{sec:results}

\providecommand{\prelim}[1]{\textcolor{blue}{#1}}
\providecommand{\tbd}{\textcolor{red}{\textbf{??}}}

\noindent
We examine qFHRR across associative retrieval, resonator factorization, path integration, and finally assess whether the storage benefits of reduced precision translate into improved efficiency on general-purpose hardware.

% ---------------------------------------------------------------- M2

\subsection{Associative Capacity}
\label{sec:res-capacity}

\noindent
\textit{Experiment.}
We store $L$ key--value pairs as a bundle of bound vectors, unbind a query key from the memory, and compare the result against a pool of 1024 candidate atoms to identify the most similar value.
% Capacity is $L_{95}$, the largest load that maintains at least 95\% top-1 accuracy, evaluated at matched dimensionality and under the 16~Kibit per-atom storage budget of Section~\ref{sec:methods}.
Capacity $L_{95}$ is the largest load that maintains at least 95\% top-1 accuracy.

\textit{Results.}
Retrieval capacity increases approximately linearly with $D$ for every format (Fig.~\ref{fig:res-tradeoff}a).
At matched $D$, reducing phase precision lowers qFHRR capacity by a factor that closely follows the predicted squared alignment $\operatorname{sinc}^{2}(\pi/K)$.
One- and two-bit qFHRR retain 0.378--0.410 and 0.777--0.825 of complex FP32 capacity, against predicted ratios of approximately 0.405 and 0.811.
From four bits onward, qFHRR remains within 3\% of FP32 at every dimensionality. 
Cartesian formats with at least four bits per stored component achieve similar capacity, and MCR performs similarly to or slightly below qFHRR at equal phase quantization.

The picture changes when storage is held constant (Fig.~\ref{fig:res-tradeoff}b), because lower phase precision permits larger dimensionality within the same budget.
One-bit qFHRR at $D=16384$ achieves $L_{95}=426.76$ and two-bit qFHRR at $D=8192$ achieves $L_{95}=427.08$.
These nearly identical capacities agree with the predicted relationship $\operatorname{sinc}^{2}(\pi/2)=\operatorname{sinc}^{2}(\pi/4)/2$.
Complex FP32 stores approximately 17 associations under the same budget, so one- and two-bit qFHRR provide roughly 25 times its retrieval capacity.
Bundling introduces a further factor.
Under the same budget, an unnormalized Cartesian int4 memory stores approximately 166 associations whereas the normalized 8-bit formats store 128--130. This suggests that projection to the unit circle discards useful information about the accumulated components.

% ---------------------------------------------------------------- M3

\subsection{Resonator Factorization}
\label{sec:res-factor}

\noindent
\textit{Experiment.}
We evaluate how quantization affects the recovery of factors from composite representations using resonator networks~\cite{frady2020resonator1,kent2020resonator2}.
Each composite $\mathbf{s}$ binds $F\in\{2,3,4\}$ factors $\mathbf{x}_f$ drawn from codebooks of $M$ candidate atoms $\mathbf{c}_{f,m}$.
At iteration $t$, the resonator unbinds its current estimates $\hat{\mathbf{x}}_j^{(t)}$ of the other factors from the composite, compares the residual with the codebook under the representation's similarity, and forms the next estimate as a rectified, weighted superposition of the candidates:
\begin{align}
\mathbf{r}_f^{(t)}
&= \mathbf{s}\odot
\bigodot_{j\neq f}
\left(\hat{\mathbf{x}}_j^{(t)}\right)^{-1},
\\
w_{f,m}^{(t)}
&= \max\!\left(
0,\operatorname{sim}(
\mathbf{r}_f^{(t)},\mathbf{c}_{f,m})
\right),
\\
\hat{\mathbf{x}}_f^{(t+1)}
&= \Pi\!\left(
\sum_{m=1}^{M}
w_{f,m}^{(t)}\mathbf{c}_{f,m}
\right),
\label{eq:resonator-update}
\end{align}
where $\Pi$ is the representation-specific projection.
All formats share this procedure and its stopping criteria, which require 2 stable iterations for convergence with limits of 100 iterations and 10 restarts.
We evaluate clean composites, corrupted composites, and composite bundles at $D\in\{1024,2048,4096\}$.
A trial is considered a success when all factors of the target composite are recovered.

\textit{Results.}
Factorization becomes harder as the number of candidate combinations grows relative to $D$. For complex FP32 at $D=2048$ and $F=3$, success falls from 100\% at $M=32$ to 36\% at $M=256$.
To compare representations, we measure paired success differences relative to complex FP32 across the 26 configurations of dimensionality, factor count, codebook size, and input condition in which FP32 achieves between 5\% and 95\% success (Fig.~\ref{fig:res-tradeoff}c).
At matched dimensionality, the reported Cartesian formats, MCR with $N\geq4$, and qFHRR with $N\geq4$ remain within two percentage points of FP32.
Reducing precision further increases the mean loss to approximately 5, 18, and 34 percentage points for three-, two-, and one-bit qFHRR, respectively.
Moreover, low-bit qFHRR reaches competitive factorization accuracy with substantially less storage (Fig.~\ref{fig:res-tradeoff}d).
% In the clean $F=3$, $M=128$ condition, two-bit qFHRR even matches complex FP32 at matched $D=2048$ (88\% versus 86\%) and exceeds it at $D=4096$ (98.5\% versus 93.5\%), whereas with bundled composites it falls below FP32 at every $D$.
% This is compatible with the 18-point mean loss in Fig.~\ref{fig:res-tradeoff}c, which averages 26 configurations of which only four favor two-bit qFHRR, all with $F=3$ and by at most 5 points.

The recovery procedure is increasingly sensitive to low phase precision, and corruption and bundling compound the difficulty.
To determine whether these failures reflect an inability to distinguish the correct factors or a limitation of iterative recovery, we compared the resonator against exhaustive search over all $M^F$ candidate tuples.
Across 22{,}980 evaluated trials, exhaustive search correctly ranks the factored tuple, yet for one-bit qFHRR at $D=1024$, $F=4$, and $M=32$ the resonator fails in all 20 trials.
Some recovery failures therefore occur even though the correct combination remains distinguishable under the evaluated similarity measure.
This distinction separates the information a quantized representation preserves from the retained factorizability.

% ---------------------------------------------------------------- M4

% Revised latter part of Results and Discussion (NICE 2027 qFHRR)
% Preserves figure/table labels and source data, but streamlines the narrative.
\subsection{Spatial Encoding and Path Integration}
\label{sec:res-spatial}

\noindent
\textit{Experiment.}
Unlike retrieval and factorization, path integration must preserve small changes in phase over repeated updates.
We encode position using fractional powers~\eqref{eq:qpow}, $\phi(x)_d=\lfloor x\tilde b_d\rceil\bmod K$, where $\tilde b_d$ is the signed index of a random base phase.
We compare per-step binding, which rounds every displacement $\delta_t$ to the stored $N$-bit resolution, with a wide accumulator that retains $w$ additional fractional bits until readout:
\begin{align}
  k_{t+1,d}
  &=\big(k_{t,d}+\lfloor\delta_t\tilde b_d\rceil\big)\bmod K,
  \label{eq:path-bind}\\
  a_{t+1,d}
  &=a_{t,d}+\lfloor 2^w\delta_t\tilde b_d\rceil,
  \label{eq:path-wide}\\
  k_{t,d}
  &=\lfloor a_{t,d}/2^w\rceil\bmod K.
  \nonumber
\end{align}
Unless otherwise stated, the base vectors, stored states, and decoding codebooks all use $N$ bits.
We test straight lines, random walks, and closed loops over 10 to 10{,}000 steps in 1D, 2D, and 3D, using $w\in\{4,8\}$ for qFHRR and MCR, and native arithmetic for the complex and Cartesian formats.
Tracking error is measured in normalized units of the kernel length scale.
Straight-line endpoints are decoded on a 1D grid of candidate positions and the 3D trefoil is decoded on a $31^3$ grid ($\pm1.5$ at spacing 0.1) refined twice around the best candidate to a resolution of 0.005.

\begin{figure}[t]
  \centering
  \includegraphics[width=0.85\columnwidth]{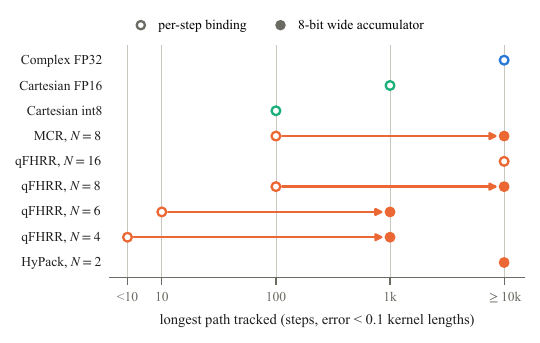}
  \caption{Longest path tracked with endpoint error below 0.1 kernel lengths ($D=1024$). Hollow: per-step rounding; filled: eight extra fractional accumulator bits. HyPack uses 16-bit bases and a two-bit stored state.}
  \label{fig:res-spatial}
\end{figure}

\textit{Results.}
The primary limitation of low-precision path integration is that small displacements can disappear before they accumulate (Fig.~\ref{fig:res-spatial}).
Since $|\tilde b_d|\leq K/2$, per-step binding leaves every component unchanged when $|\delta_t|<2^{-N}$.
For a unit-length line divided into $n$ updates, this predicts a complete loss of motion when $n>2^N$.
Consistent with this limit, eight-bit qFHRR tracks 100 steps but fails by 1{,}000, with similar behavior for Cartesian int8 and MCR.
Cartesian FP16 also exceeds the tracking threshold beyond 1{,}000 steps, whereas 16-bit qFHRR and complex FP32 remain accurate through 10{,}000.
Increasing $D$ from 256 to 4096 can reduce interference, but it does not restore phase increments discarded during quantized updates.

\begin{figure}[t]
  \centering
  \includegraphics[width=\columnwidth]{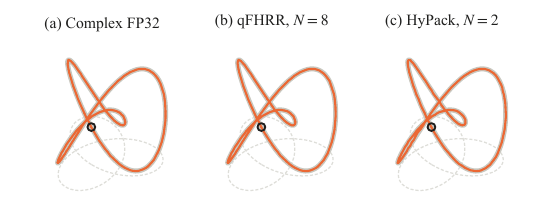}
  \caption{Decoded 3D trefoil (1000 steps, $D=1024$, one seed): (a) complex FP32 with per-step binding; (b) eight-bit qFHRR with $w=8$; and (c) HyPack with a 16-bit base, wide accumulator, and two-bit readout. Maximum errors are 0.004, 0.004, and 0.018 kernel lengths (decoder resolution 0.005). Gray: reference; dashed light gray: its projection onto the floor; orange: decoded every five steps; circle: start. Each format uses its own codebook and similarity.}
  \label{fig:res-paths}
\end{figure}

Retaining fractional precision during accumulation addresses this failure mode.
The wide accumulator reduces the zero-update threshold to $2^{-(N+w)}$ and allows successive displacements to contribute to the state before it is quantized for storage.
% Closed trajectories make the improvement particularly clear because endpoint error alone can hide a failure to move.
Across 20 seeds on a 10{,}000-step 3D trefoil at $D=1024$, per-step rounding never changes the eight-bit qFHRR state, so the decoded position stays at the start with a maximum error of 1.46 kernel lengths.
With eight fractional accumulator bits, the maximum error falls to approximately 0.004 kernel lengths and is comparable to complex FP32 and 16-bit qFHRR. Fig.~\ref{fig:res-paths} shows a single-seed 1{,}000-step example.
These results separate the precision needed to accumulate motion, governed by $N+w$, from the residual error of the quantized readout.

The base vectors introduce a separate constraint.
Fractional power encoding represents position through sinusoids whose frequencies are set by the base phases.
Quantizing those phases to $N$ bits places the frequencies on a discrete grid and makes the ideal encoding periodic with period $2^N$ kernel lengths along each axis.
It also limits the number of distinct frequency combinations: in 3D an $N$-bit base provides at most $2^{3N}$ unique combinations, regardless of $D$.
For $N=2$, this upper bound is only 64, and the native two-bit base exhibits at least 0.25 kernel lengths of trefoil error even with a wide accumulator.
Quantizing the \emph{stored state} after accumulation is different: it retains the base frequencies, and its componentwise readout error partially averages across dimensions when the phases occupy the available bins.

This distinction motivates HyPack's mixed-precision design.
HyPack uses 16-bit base phases and a wide integer accumulator to preserve motion, but stores and compares the resulting state using only two-bit phase indices.
It tracks the 10{,}000-step straight line and closely follows the 3D trefoil with a two-bit readout (Figs.~\ref{fig:res-spatial} and~\ref{fig:res-paths}).
For the 1{,}000-step trefoil shown in Fig.~\ref{fig:res-paths}, its maximum decoded error is 0.018 kernel lengths, compared with 0.004 for complex FP32 and eight-bit qFHRR.

\begin{table*}[t]
  \centering
  \caption{Accuracy-matched hardware performance ($D^\ast=4096$ for retrieval; $D^\ast=1024$ for path integration). Fused rows use one C++/OpenMP or Triton kernel; others use PyTorch. Stored/State: stored bits per dimension and, for path integration, accumulator-state bits per dimension. Codebook: 1024 candidate values for retrieval; for path integration, the decoder's 701 candidate positions ($701\times D$). Time is the median over 10 tasks; energy is the mean above idle; retrieval time and energy are per query within a 128-query batch. Bold denotes the minimum time, energy, and codebook size per workload. Orin: Jetson Orin Nano Super (6-core Cortex-A78AE, 1024-core Ampere GPU) in the 25~W mode with CPU, GPU, and memory clocks held at 1.73, 1.02, and 3.2~GHz; energy is the shared CPU+GPU supply rail (INA3221).}
  \label{tab:res-hw}
  \setlength{\tabcolsep}{4pt}
  \scriptsize
  \setlength{\aboverulesep}{0pt}\setlength{\belowrulesep}{0pt}\renewcommand{\arraystretch}{1.1}
  \begin{tabular}{@{}llc|S[table-format=2.0]S[table-format=2.0, table-text-alignment=right]|S[table-format=2.2]|S[table-format=5.1]S[table-format=6.0]|S[table-format=5.2]S[table-format=6.1]|S[table-format=5.1]S[table-format=6.0]|S[table-format=5.1]S[table-format=5.1]@{}}
    \toprule
    & & & \multicolumn{2}{c|}{Bits/dim} & {Codebook} & \multicolumn{2}{c|}{3700X CPU} & \multicolumn{2}{c|}{RTX 3090 GPU} & \multicolumn{2}{c|}{Orin CPU} & \multicolumn{2}{c}{Orin GPU} \\
    \cmidrule(lr){4-5}\cmidrule(lr){6-6}\cmidrule(lr){7-8}\cmidrule(lr){9-10}\cmidrule(lr){11-12}\cmidrule(lr){13-14}
    Workload & Format & Fused & {Stored} & {State} & {MB} & {$\mu$s} & {$\mu$J} & {$\mu$s} & {$\mu$J} & {$\mu$s} & {$\mu$J} & {$\mu$s} & {$\mu$J} \\
    \midrule
    Retrieval
      & Complex FP32 &  & 64 & {---} & 33.55 & 69.3 & 5000 & 2.89 & 779 & 256.3 & 1280 & 30.6 & 302 \\
      & Complex FP32 & \checkmark & 64 & {---} & 33.55 & 35.1 & 2740 & 8.33 & 1630 & 1740.0 & 6710 & 46.5 & 192 \\
      & Cartesian FP16 &  & 32 & {---} & 16.78 & 144.8 & 11300 & 2.31 & 329 & 26873.7 & 144000 & 8.0 & 44.0 \\
      & Cartesian FP16 & \checkmark & 32 & {---} & 16.78 & 30.0 & 2350 & 1.01 & 265 & 1718.4 & 7080 & \bfseries 3.8 & 29.0 \\
      & Cartesian int8 &  & 16 & {---} & 8.39 & 3210.4 & 177000 & 3.33 & 677 & 16652.4 & 43000 & 17.9 & 104 \\
      & Cartesian int4 &  & 8 & {---} & 4.19 & 3044.6 & 173000 & 3.32 & 667 & 16907.4 & 43800 & 18.0 & 102 \\
      & qFHRR, $N=4$ &  & 4 & {---} & 2.10 & 263.7 & 12500 & 4.84 & 1550 & 316.7 & 1190 & 41.6 & 283 \\
      & qFHRR, $N=2$ &  & 2 & {---} & \bfseries 1.05 & 252.3 & 11800 & 4.85 & 1510 & 319.8 & 1100 & 41.6 & 255 \\
      & MCR, $N=4$ &  & 4 & {---} & 2.10 & 26903.4 & 917000 & 290.71 & 90100 & 9744.0 & 26400 & 2643.8 & 10400 \\
      & HyPack, $N=2$ &  & 2 & {---} & \bfseries 1.05 & \bfseries 9.1 & \bfseries 549 & \bfseries 0.73 & \bfseries 99.9 & \bfseries 37.1 & \bfseries 153 & 4.3 & \bfseries 19.9 \\
    \midrule
    Path integration
      & Complex FP32 &  & 64 & 64 & 5.74 & 1906.9 & 41000 & 6031.96 & 625000 & 4722.8 & 3720 & 32506.3 & 27000 \\
      & Complex FP32 & \checkmark & 64 & 64 & 5.74 & 42.1 & \bfseries 1550 & 64.68 & 6420 & \bfseries 132.5 & \bfseries 350 & 194.5 & \bfseries 195 \\
      & Cartesian FP16 (wide) & \checkmark & 32 & 64 & 2.87 & 42.4 & 1610 & \bfseries 62.14 & \bfseries 6130 & 652.3 & 1430 & \bfseries 180.6 & 202 \\
      & qFHRR, $N=8$ (wide) &  & 8 & 16 & 0.72 & 2123.3 & 51100 & 5721.92 & 609000 & 2759.0 & 2610 & 24751.2 & 20200 \\
      & HyPack, $N=2$ (wide) & \checkmark & 2 & 24 & \bfseries 0.18 & \bfseries 30.3 & 1700 & 76.57 & 6690 & 157.4 & 463 & 220.1 & 263 \\
    \bottomrule
  \end{tabular}
\end{table*}

\subsection{Hardware Efficiency}
\label{sec:res-hw}

\noindent
We next examine whether compact phase representations improve execution time and energy.
Table~\ref{tab:res-hw} compares accuracy-matched retrieval and path-integration workloads on the Ryzen 7 3700X CPU, the RTX 3090 GPU, and the CPU and GPU of a Jetson Orin Nano Super.
Fused complex FP32 and Cartesian FP16 kernels separate the effect of kernel fusion from that of computing directly on packed phases.

\textit{Retrieval.}
The framework benchmarks first show why reduced storage does not guarantee faster inference.
At a load of 128 stored pairs and 1024 candidates, every eligible format requires $D^\ast=4096$ to meet the accuracy threshold.
Despite its smaller representation, decoded qFHRR is 3.6--3.8$\times$ slower than complex FP32 on the CPU and approximately 1.7$\times$ slower on the GPU.
The bottleneck is similarity search rather than binding, where quantized representations must either convert the candidate codebook values to Cartesian values before performing a floating-point matrix multiplication or leverage inefficient LUT arithmetic.
Other compact framework implementations are similarly constrained by their arithmetic, including Cartesian int8/int4 and MCR on the CPU.

HyPack avoids the conversion bottleneck by keeping unbinding and similarity on Gray-coded bit planes through XOR and population count.
It produces the same scores as decoded two-bit qFHRR while reducing the 1024-candidate codebook from 33.55~MB in complex FP32 to 1.05~MB.
Against the fastest fused baseline on the CPU, Cartesian FP16, HyPack is 3.3$\times$ faster and uses 4.3$\times$ less energy per query.
On the RTX 3090, where each HyPack call runs as one Triton launch, HyPack is 1.4$\times$ faster than the fused FP16 tensor-core kernel and uses 2.6$\times$ less energy.
% Fusing the key gather and packed unbind into the popcount kernel is what secures this margin: measured in the same run, it lowers the latency of the query by 2.1$\times$ and its energy by 1.9$\times$ relative to issuing the unbind as separate framework operations, a version that was 1.5$\times$ slower than the fused FP16 kernel.
% The margin is smaller on the GPU than on the CPU: packed phase arithmetic is advantageous when it avoids expensive decoding and launch overhead, but tensor-core floating-point kernels remain highly competitive on GPUs.

\textit{Path integration.}
For path integration, all eligible configurations use $D^\ast=1024$.
In the baseline framework implementations, eight-bit qFHRR with a wide accumulator performs similarly to complex FP32.
Therefore, changing the stored precision alone does not remove the overhead of a step-by-step loop.
Fusing the updates into one kernel has a much larger effect, accelerating complex FP32 by 45$\times$ on the CPU and 93$\times$ on the GPU.
A similarly fused Cartesian FP16 implementation, with its accumulation state retained in FP32 until readout, performs comparably.

HyPack performs these updates with 16-bit base phases and eight fractional accumulator bits, a 24-bit state held in an int32 container, before producing a packed two-bit readout.
Its fused integration kernel takes 30.3~$\mu$s on the CPU, compared with 42.1~$\mu$s for fused complex FP32, but uses slightly more energy (1700 versus 1550~$\mu$J).
On the GPU, HyPack takes 76.57~$\mu$s versus 64.68~$\mu$s for complex FP32, at similar energy.
The integer update therefore offers no consistent execution advantage across the two devices.

The larger advantage appears when the integrated position must be decoded.
Retrieving a position requires scoring the final state against candidate position encodings.
For HyPack, this is an XOR--popcount comparison against a 0.18~MB packed codebook, rather than a complex similarity search over 5.74~MB of FP32 values.
On the CPU, decoding takes 10~$\mu$s and 432~$\mu$J with HyPack, compared with 135~$\mu$s and 9120~$\mu$J for complex FP32.
Including both integration and position decoding, HyPack takes 40~$\mu$s and 2130~$\mu$J, versus 177~$\mu$s and 10{,}700~$\mu$J for fused complex FP32: a 4.4$\times$ latency improvement with approximately 5.0$\times$ less energy.
On the GPU, the single-query decoder is more strongly affected by launch overhead where HyPack's 100~$\mu$s end-to-end time is approximately 1.3$\times$ faster than fused complex FP32 but remains within 8\% of fused Cartesian FP16.
Table~\ref{tab:res-hw} times integration only.

\textit{Jetson Orin Nano Super.}
The embedded device reproduces this pattern.
On its CPU, the fused C++ kernels run their generic path without SIMD intrinsics, so fused complex FP32 is 6.8$\times$ slower than the framework GEMM and the fastest baseline is framework complex FP32 at 256~$\mu$s and 1280~$\mu$J per query.
HyPack answers the same query in 37~$\mu$s with 153~$\mu$J, 6.9$\times$ faster with 8.3$\times$ less energy, whereas the decoded qFHRR formats are 1.2$\times$ slower than complex FP32 at 0.9$\times$ the energy.
On the Orin GPU, the HyPack query takes 4.3~$\mu$s and 19.9~$\mu$J, 7.1$\times$ faster than framework complex FP32 with 15$\times$ less energy, but the fused FP16 tensor-core kernel remains 1.1$\times$ faster at 1.5$\times$ the energy.
Integration alone favors the fused floating-point kernels on both Orin processors, but decoding reverses the order: HyPack takes 187~$\mu$s and 579~$\mu$J on the CPU against 447~$\mu$s and 1680~$\mu$J for fused complex FP32 (2.4$\times$ faster with 2.9$\times$ less energy), and 308~$\mu$s and 435~$\mu$J on the GPU against 544~$\mu$s and 1520~$\mu$J (1.8$\times$ and 3.5$\times$).

Moreover, HyPack's compact readout retains an accuracy cost that is not captured in the latency analysis.
On the trefoil in Fig.~\ref{fig:res-paths}, HyPack's maximum decoded position error is 0.018 kernel lengths, compared with 0.004 for the eight-bit readout.
Taken together, the hardware results show that HyPack's principal benefit is not universally faster integration, but the ability to retain accurate updates while keeping subsequent similarity searches in a compact, natively supported representation.

% ---------------------------------------------------------------- synthesis
\subsection{Discussion}
\label{sec:res-discussion}

\noindent
The precision FHRR requires depends on the task: added dimensions compensate for coarse phases in retrieval but not in iterative factorization, and path integration needs precise accumulation but only a coarse stored state.
Storage and intermediate precision should therefore be chosen independently.

These results are related to spiking phasor networks, which encode a phase as a spike time within an oscillation cycle~\cite{frady2019rhythmic,orchard2023spikingphasor,renner2024neuromorphic,orchard2024spikingphasors}.
An $N$-bit phase corresponds to $2^N$ spike-timing bins per cycle, so, without testing spiking dynamics directly, our findings suggest that associative retrieval can tolerate one or two bits of timing resolution when the population is enlarged, that resonator networks need about four bits, or 16 bins per cycle, at matched $D$, and that path integration needs fine integration ($N+w$ bits) but only a coarse readout.
The last case points to integrating displacements in a finer internal state, such as membrane or dendritic potentials, while emitting coarsely timed spikes.

On the hardware side, reduced precision pays off only when the processor can operate on the compact representation directly, as HyPack's bitwise similarity does on both desktop and embedded platforms.

\section{Conclusion}

\noindent
We investigated how much numerical precision FHRR requires by comparing phase-based and Cartesian representations across associative retrieval, resonator factorization, and spatial path integration.
At matched dimensionality, the reduction in associative retrieval capacity follows the predicted relationship $\operatorname{sinc}^{2}(\pi/2^{N})$.
Under a fixed storage budget, however, increasing dimensionality can more than compensate for this loss.
One- and two-bit qFHRR support about $26\times$ more key--value associations than complex FP32 under a 16\,Kibit storage budget.
Resonator factorization is more sensitive to reduced precision because of its iterative recovery procedure, while path integration requires sufficient precision to preserve small phase increments.
Thus, the precision needed to store a representation can differ from that needed to manipulate it.

To exploit this distinction, we introduced HyPack, a hybrid-precision implementation that combines packed two-bit phase representations with native bitwise operations and wider integer accumulation.
Compared with the fastest fused floating-point baselines, HyPack accelerates associative retrieval by $3.3\times$ on a desktop CPU and $1.4\times$ on a desktop GPU, while reducing energy consumption by factors of $4.3$ and $2.6$, respectively.
Although its integer path integration kernel does not consistently outperform floating-point alternatives, its compact readout and codebook substantially reduce the cost of position decoding.
This produces a $4.4\times$ speedup and a fivefold reduction in energy consumption on the desktop CPU.
Similar trends emerge on the Jetson Orin Nano Super, demonstrating the benefits and limitations of packed phase arithmetic across desktop and embedded processors.

Together, these findings suggest that FHRR should not be constrained to a uniform numerical precision.
Instead, representation precision, dimensionality, and intermediate arithmetic should be selected according to the computational task and the capabilities of the underlying hardware.
Future work will investigate these tradeoffs on dedicated digital accelerators and spike-based neuromorphic architectures, where native phase-based computation may provide further improvements in efficiency.
\section*{Acknowledgments}

\noindent
This work was supported in part by the National Science Foundation (NSF) under Grants 2539714 and 2319619.

\bibliographystyle{ieeetr}
\bibliography{bibliography}

\end{document}